# Astrometric accuracy during the past 2000 years

*Erik Høg, Niels Bohr Institute, Copenhagen*

ABSTRACT: The great development of astrometric accuracy since the observations by Hipparchus about 150 BC was documented in 2008 in the first version of the present report. This report was updated in Høg (2017d), e.g. with recent information on the catalogues before 1800 AD. The development has often been displayed in diagrams showing the accuracy versus time. A new simplified diagram is provided in the pesent report as Fig.2 and in a .png file and this information will presumably be the main interest for some readers.

## 1 Introduction

The present report shall briefly present the accuracy of observed positions of stars during the past 2000 years. More information is available in Høg (2017d) hereafter called EH17d which includes an appendix from 2008. Historical information and explanations are partly taken over from this report as needed for a proper understanding of the development.

The evolution has often been displayed in diagrams, showing the accuracy versus time. These diagrams have at least one thing in common, the improvement by many powers of ten from the half degree errors of Hipparchus, the Greek father of astronomy, to one milliarcsec (mas or millisecond of arc) for the diagrams including the Hipparcos Catalogue. But Tycho Brahe and Flamsteed are the only other sources always included, though with quite different numbers. Other differences are pointed out in EH17d. I will here present a recommended simplified diagram of astrometric accuracy primarily of positions.

The diagrams in Figs. 2, 3, 8, 10 in the appendix from 2008 show a line joining the points which does not give a clear meaning in a plot of discrete points. A variant of Fig. 8 was also shown at a recent conference. These diagrams give the impression of a smooth, gradual improvement over all the centuries, including the last 500 years. This obscures the historically interesting fact that jumps can be clearly seen in Fig. 1 from 2016 and already in Fig. 1b of EH17d from 2008. A *'jump'* means a *big improvement within a very short time* as the result of great investment of material resources and intellectual efforts.

The largest jumps have been obtained in modern times by space astrometry. The Hipparcos satellite launched by ESA in 1989 gave a factor 100 over the contemporary accuracy of positions obtained from the ground. The Gaia satellite mission launched in 2013 by ESA has by 2019 yielded a further factor 100 over Hipparcos but for many more stars as was expected.

A jump in accuracy was obtained more than 400 years ago by the Landgrave in Kassel and by Tycho Brahe before 1590 when they both measured positions ten times more accurately than Hipparchus/Ptolemy and Ulugh Beg. New information about the Landgrave only reached me in 2015 although it was published by Jürgen Hamel first in 1998 and later in Hamel (2002). I am grateful to Andreas Schrimpf for lending me that book at my visit to Marburg in November 2015.

The following section contains recommended diagrams of astrometric accuracy.

## 2 Recommended diagrams

Figure 1 is a diagram of the development of astrometric accuracy with time prepared in 2016 for the report EH17d. It is called Høg-2016 since, for convenience, a diagram is designated by "name-year". It is available in the report at a link to a .png-file.

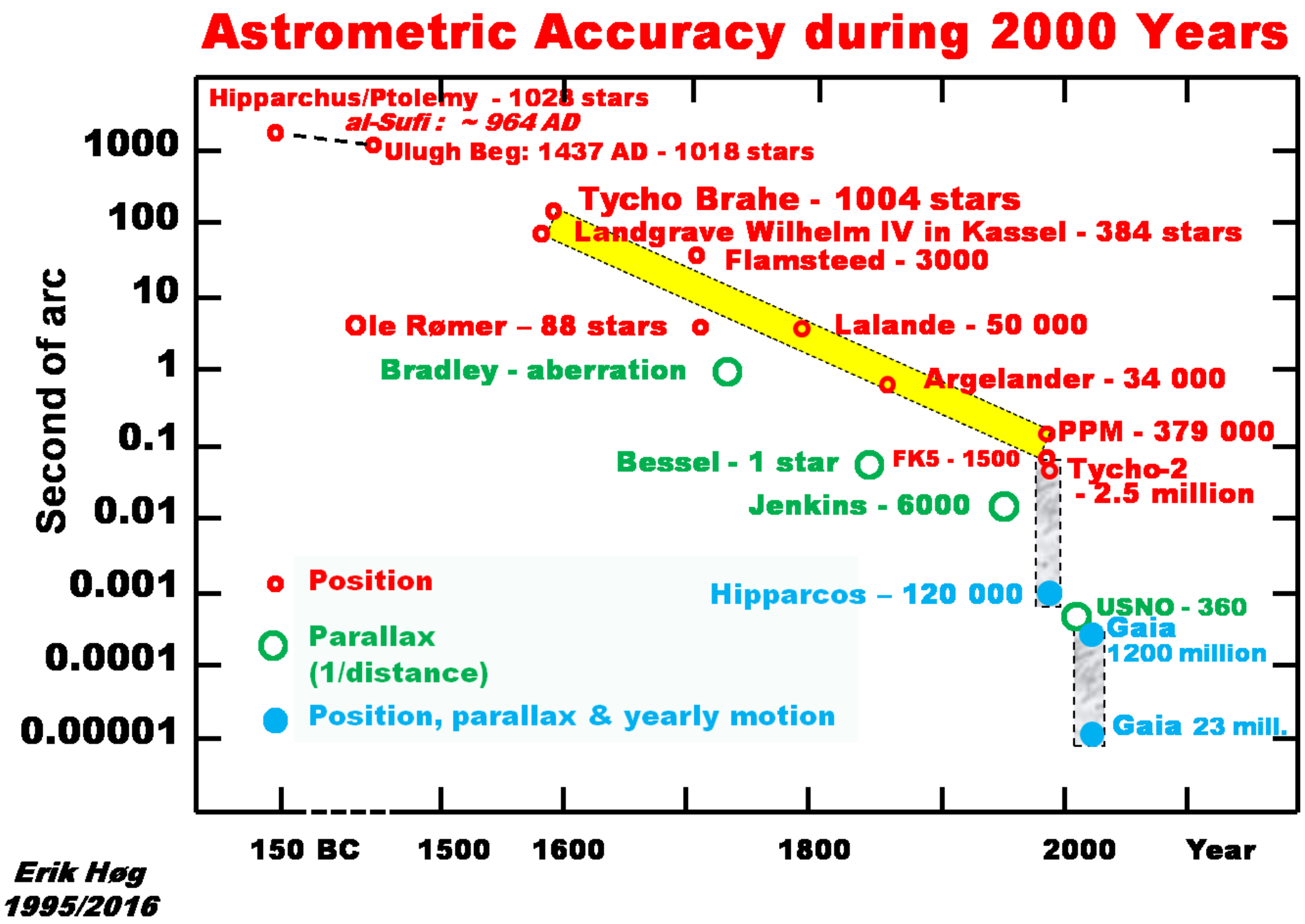

**Fig. 1. Astrometric accuracy during 2000 years: Høg-2016.** The accuracy was greatly improved shortly before 1600 AD by the Landgrave in Kassel and by Tycho Brahe in Denmark. The following 400 years brought even larger but much more gradual improvement before space techniques with the Hipparcos satellite started a new era of astrometry.

The points are placed at the mean observation epoch, except the compilation catalogues FK5, PPM, and Jenkins which are placed at the year of publication and with the accuracy of the positions in FK5 and PPM in that year. The circles refer to "positions" and "parallaxes" and we want to show median values of the standard errors in each catalogue, representative for the bulk of stars in a catalogue. It has been suggested to include more information on the most accurate stars in each catalogue, but the diagram would be more complicated and it would be very difficult to collect the information and to present it well in a single graph.

One of the points has been included for historical reasons but does not represent star positions. This point gives credit to the Persian astronomer Abd ar-Rahman al-Sufi (or as-Sufi or Azophi). He provided magnitudes of many stars in the catalogue of Ptolemy and he thus represents here the important muslim astronomers before Ulugh Beg. The points for ROEMER and SIM were included in the 2008-diagram to represent two important proposals for astrometry from space, but they were omitted in 2016.

A new simpler diagram omitting the ground based parallaxes is shown in Fig. 2.

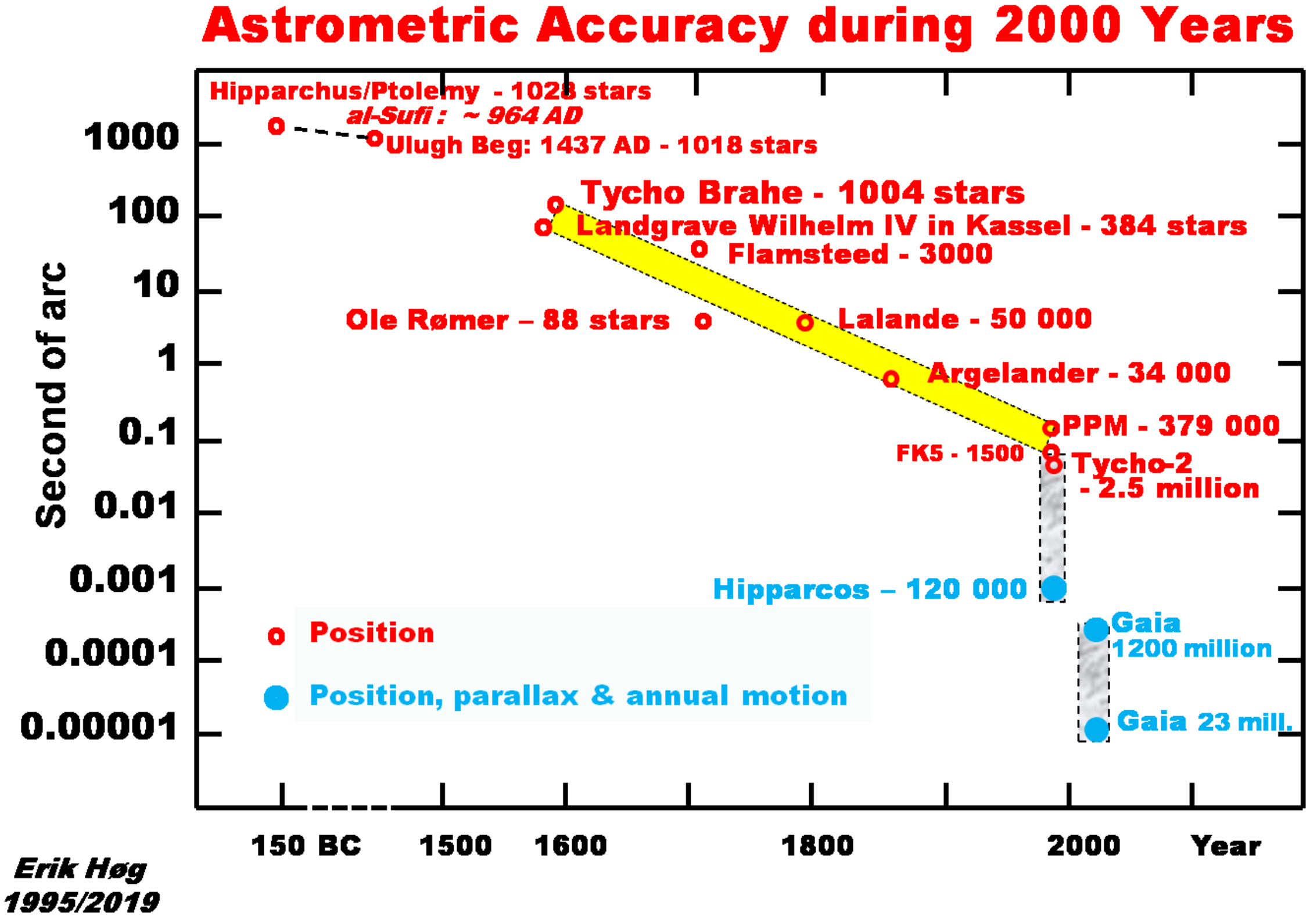

**Figure 2. Astrometric accuracy during 2000 years: Høg-2019.** The accuracy was greatly improved shortly before 1600 AD by the Landgrave in Kassel and by Tycho Brahe in Denmark. The following 400 years brought even larger but much more gradual improvement before space techniques with the Hipparcos satellite started a new era of astrometry. - This is the same explanation as in Fig. 1 but omission of the points for parallaxes makes it simpler and more suited for most presentations. For anybody to use, it is available at a link to a .png-file at Høg (2019).

**Acknowledgements in EH17d:** I am grateful to W.F. van Altena, F. Arenou, P. Brosche, J. de Bruijne, T. Corbin, D.W. Evans, D.W. Hughes, P. Høyer, J. Kovalevsky, J. Lequeux, F. Mignard, M.A.C. Perryman, H. Pedersen, A. Schrimpf, C. Turon, R. van Gent, F. Verbunt, A. Wittmann and N. Zacharias for sending information without which the present report could not have been written and/or for comments to previous versions of the report. The documentation has been collected with much support from colleagues, but I must take the responsibility for imperfections which may still be present in spite of considerable efforts on my part.

# 3 References

All links to my dropbox had become wrong in March 2017 but are now replaced by links to my website.

Here follow a few references especially to some of my reports on accuracy. More references are given in Høg (2017d) = EH17d.